\documentclass[12pt]{article}  

\begin{document}

\title{A possible scenario for thermally activated avalanches in type-II superconductors}
\author{R. Mulet, R. Cruz\thanks{Present address: ICA1, University of Stuttgart}  ~and E. Altshuler\\ 
 Superconductivity Laboratory, Physics Faculty-IMRE,\\ University 
of Havana\\ La Habana 10400, Cuba\\ } 
\date{\today}  


 \maketitle 

 \begin{abstract} 
Using a simple cellular automaton with stochastic rules we show the possible emergence of thermally activated avalanches (power law distributed) in type-II superconductors. Scaling relations between the exponents characterizing these distributions and those obtained from field driven experiments are derived and proved through simulations. It is also shown that the conditions for the appearance of these avalanches  are independent of the pinning mechanism. The relevance of our simulations for recently reported experimental results is also outlined.


\end{abstract}

\section{Introduction}

  Magnetic flux penetrates type-II superconductors 
above a certain critical field $H_{c1}$ in the form of vortices. 
The interaction of these vortices with the pinning centers  produces 
a magnetic flux profile inside the superconductor with a slope
 proportional to the critical current density inside the sample, 
$j_c$, defined as  the maximum current density the material
 supports without  dissipation, a situation which is accounted 
 by the so called Bean's Critical State Model \cite{Bean_64}. 
This picture, as de Gennes fristly noted \cite{deGennes_66}, 
is very similar to the case of sandpiles, where a constant slope 
appears in the pile resulting from the competition between gravity 
and the friction between grains.  

In 1987, Bak 
et al \cite{Bak_87,Bak_88} proposed a theory 
-- now known as Self Organized Criticallity theory 
(SOC) -- to explain the  existence of self similar
 structures in Nature.  Since then,  SOC has been used to
 interpret the dynamics of many size avalanches in 
sandpiles\cite{Held_90}, earthquakes\cite{Carlson_89}, evolution\cite{Bak_89}, 
and other phenomena, see~\cite{Bak_book} for a general review. 

The ocurrence of self-organized criticallity was soon searched also in superconductors where field driven experiments have been designed \cite{Field_95,Nowak_pp,Mailly_pp} and many numerical simulations developed \cite{Pla_91,
Richardson_94,Bassler_98,Cruz_99,Mholer_99,Mulet_97}, unfortunately without conclusive answers.

Superconductors differ from most systems exhibiting SOC by the relevant role played by the temperature. The temperature causes the relaxation of the critical state leading to a nearly logarithmic magnetization decay, $m(t) \sim ln (t)$ \cite{Malozemoff}. In the early 90's many researchers tried to relate the role played by the temperature to the existence of many size avalanches in relaxation experiments \cite{Vinokur_91,Schnack_92,Vinokur_92,Tang_93,Wang_93}. 
However in 1995 Bonabeau and Lederer \cite{Bonabeau_95,Bonabeau_96}, 
approximately solved the difusion equation for the magnetic field inside a superconducting slab and demostrated that
 (within the usually accesible time scales in the experiments) it is impossible to determine the existence of thermally activated avalanches by classical magnetic relaxation measurements, i.e by the study of the decay of the mean value of the magnetization in the sample\cite{Malozemoff,Wang_93}.

In 1998 Aegerter~\cite{Aegerter_98} studied the magnetic relaxation of a Bismuth single crystal but instead of  the usually measured mean value of $m(t)$ \cite{Malozemoff} he put attention to
 the fluctuations during the decay of the magnetization and showed evidences of power law distributed thermally activated avalanches. 

In this work we develop a simple scenario able to account for the existence of these thermally activated avalanches. This does not mean we claim for the existence of SOC during the relaxation of the magnetization. SOC is well defined only for a system in a marginal stationary state and it is not the case for the vortex lattice in the presence of thermal activation. What we are claimining is that, because of the complex interaction between vortices, the pinning centers and the temperature, many size avalanches of moving vortices may produce the relaxation of the critical state as previously determined in reference~\cite{Aegerter_98}.

The remaining of the paper is organized as follows. In the next section we describe the Cellular Automaton used in our simulations. In section 3 we present and discuss the numerical results. Then, section 4 is devoted to the study of the scaling relations between our distributions and those usually obtained in field driven experiments. In section 5 we outlined some conditions needed for the ocurrence of many sizes thermally activated avalanches and finally in section 6 the conclusions are given.

\section{The model}

While the use  of ``real'' forces between vortices
in molecular dynamics simulations~\cite{Pla_91,Richardson_94} better resemble the experimental situation than simple cellular automata, they are by far more time consumming and it is an important drawback of the method, specially when we are looking for critical exponents or when we introduce the effects of the temperature in the system. 

Recently, to neglects these problems Bassler and Paczuski \cite{Bassler_98} introduced a simple cellular automaton to study the behaviour of the vortex lattice in type-II superconductors.
This cellular automaton  avoids part of the relevant physics of the
vortex lattice such as the variation of the pinning strenght with  the increasing field, the possible mistmach between the vortex lattice  and the pinning centers, the elasticity of the vortex lattice, etc. However, it contains the interaction between vortices and with the pinning centers, the long range order of the vortex interaction, first by the introduction of the parameter $r$ (see below), but also implicity assuming that each lattice cell contains more than one vortex. In addition it is able to predict the self organization of the lattice in a critical state characterized by power law distributed avalanches\cite{Bassler_98,Cruz_99} and the irreversibility of the magnetization. Then, aimed at describing the influence of the temperature on this critical state we adopt this model with some modifications.

The cellular automata consists in a $2D$ honeycomb lattice, where
 each site  is characterized by the number of vortices on it $m(x)$,
 and by its  pinning strength  $V(x)$ equal to 0 with probability 
$p$, and to $q$ with probability $1-p$. The force acting on a vortex
 at site $x$ in the direction  to site $y$ is calculated as: 

 \begin{eqnarray} F_{x \rightarrow y}= -V(x)+V(y) +(m(x)-m(y)-1)
 \nonumber\\ +r[m(x_1)+m(x_2)-m(y_1)-m(y_2)] \label{eq:force}
 \end{eqnarray} 

 \noindent where $x_1$ and $x_2$ are the nearest neighbors of 
$x$ (other than $y$) while  $y_1$ and $y_2$ are the nearest neighbors 
of $y$ (other than $x$) and $r$ is a measure of their 
contribution to the total force on the 
vortex $x$ $(0<r<1)$.
A vortex in the site $x$, moves to its neighbor  site $y$ if the
 force acting on it in that direction is greater than zero.  If
 the force in more than one direction is greater than zero, then
 one of them is chosen at random~\cite{Bassler_98,Cruz_99,Mholer_99}.  

To introduce the effect of the
 temperature we assumed that sites where the forces are lower than
 zero still have a probability of motion given by:  

\begin{equation} P_{x \rightarrow y} \sim \exp(-U(j)/kT)
\label{eq:probability} \end{equation}  

\noindent where $k$ is the Boltzman's constant and $T$ is the temperature. The current,
$j$, was locally  calculated using the gradient of $m(x)$  and 
$U(j)$ represents  different pinning barriers proposed in the literature 
$U(j) = U_o j_c/j$, $U(j) = U_o \ln(j_c/j)$ and 
$U(j) = U_o (1-j/j_c)$ \cite{Blatter_94}. 

An avalanche starts by randomly choosing a lattice site, 
and calculating  (\ref{eq:probability}). If it is smaller
than a random number the procedure is repeated, else the 
vortex moves perturbing its neighbours. Then,  the direction 
of motion of the new unstable vortices is calculated 
using (\ref{eq:force}). At this point,  all the sites 
are updated in parallel until no more unstable sites persist.
The avalanche size is defined as the number of topplings corresponding to the thermal activation of one vortex while the avalanche duration is defined as the number
of updatings necessary to complete one avalanche.

 In all the cases the procedure was repeated for $10^{4}\,m.c.s$, 
were one $m.c.s$ was  defined by the $L^2$ calculation of 
(\ref{eq:probability}),  and lattices up to $L=200$ were used. 
 The initial configuration was obtained slowly adding 
vortices to the system (at $T=0$) until a critical slope was reached
 \cite{Bassler_98}. The boundaries  ``parallel to the net vortex motion''
were assumed periodic  while the other two were fixed to mimic the applied external field.

The magnetization, $M$ was calculated as the mean magnetic field inside the sample minus the external applied field~\cite{Mulet_97}, i.e.,

\begin{equation}
 M = \sum^{i=L}_{i=0} B(i)-H
\label{eq:magn}
\end{equation}

\noindent where $H$ is the field, i.e. the number of vortices,  at the borders of the lattice.

\section{Numerical results}

Figure 1 shows typical relaxation curves obtained for systems
 of different sizes using
a vortex glass-like potential $U(j)\sim j_{c}/j$ \cite{Blatter_94} and the algorithm described above. In Figure 2 is represented the relaxation curve for 
a system with $T \sim \infty$, it means when  we disregard the avalanche-like behavior previosly explained avoiding the calculation of equation~(\ref{eq:force}) after a thermal jump.

In both figures (see also the inset) three regimes are 
present, a plateau, then a logarithmic relaxation, and finally 
another plateau due to finite size effects. Only the time scales
 for these regimes are different, but this is irrelevant from 
the experimental point of view.  
So, as already noted before \cite{Bonabeau_95,Bonabeau_96} our results suggest that it is not possible 
to  decide about the existence or not of thermally activated 
vortex avalanches  from ``simple thermodynamic magnetic'' relaxation measurements.
Other pinning potential as well as different $Uo/kT$ relations were used \cite{Blatter_94} and no fundamental differences with the previous results were obtained.

 Figure 3 and 4 represent the integrated\footnote{The meaning of this name will be clarified below} distribution of avalanche sizes,
 $D_{int}(s)$, and the integrated  distribution of avalanche times, 
$D_{int}(t)$  obtained using a classical Anderson-Kim 
potential, $U(j)=U_o (1-j/j_c)$~\cite{Anderson_64,Blatter_94},  for a system with $L=200$ and $U_o/kT=10$, as before, other 
pinning potentials were also used, resulting in a similar
 behavior. These distributions were obtained using the avalanche sizes and times (defined in section 2) obtained during all the relaxation process.

As figure 3 an 4 clearly show, many size avalanches 
emerge. It {\em does not} mean the system is critical, instead it is relaxing from   a critical state to its corresponding thermodynamic equilibrium. What we are showing is that this relaxation could proceed by means of many size avalanches in accordance with recent experimental results \cite{Aegerter_98}.
However, somehow more surprisingly, we will show in the next section that the exponents characterizing these distributions are related through simple scaling relations to the exponents derived in the context of SOC for systems in a critical state \cite{Bassler_98,Cruz_99}. 

Considering that $D_{int}(s)$ follows a  power law:

\begin{equation}
 D_{int}(s) \sim s^{-\tau_{n}}
\label{eq:power_law1}
\end{equation}

\noindent the estimated exponent form figure 3 was
$\tau_{n}=2.70\pm0.1$ (different form the $\tau=1.63$ obtained 
in references
 \cite{Bassler_98} and \cite{Cruz_99} for a field driven experiment)
 and, assuming $D_{int}(t) \sim t^{-\tau_{tn}}$
for the integrated distribution of avalanche times, we got  from figure 4, 
$\tau_{tn}=4.0\pm0.2$. 

It is worth to mention here that the exponent $\tau_{n}$ was also 
experimentally determined in reference \cite{Aegerter_98} and 
reported as $\tau_{n}=2.0$, lower than our value. This divergence can be explained since  figures 3 and 4 represent the distribution of avalanches obtained for all the relaxation process, i.e starting at the critical state and finishing at equilibrium, a situation imposible to account for in real experimental situations. 

To determine the distribution of avalanche sizes, using just part of the relaxation curves, gives different numerical estimates for $\tau_{n}$ and $\tau_{tn}$ as is evident from figure 5. In fact, figure 5 represents five avalanche size distributions, $P(s)$, obtained for different time intervals of  the relaxation curve, from the upper to the lower curve, $t=1-10$, $t=11-100$, $t=101-1000$, $t=1001-10000$ and $t=10001-100000$ m.c.s, which superposition corresponds to the full relaxation of the system (see figure 1). The straight line represents the integrated distribution of avalanche sizes, $D_{int}(t)$, obtained in figure 3, $\tau=2.7$.
 Then, from  the figure we can conclude that can be predicted different exponents depending on the range of time measured. For low enough times, the exponent is lower than that associate to $D_{int}$, while for large times a peaked distribution is obtained with only very small avalanches. 

Another source for discrepancies between our numerical estimates and experimental situations comes from the change of regimes of relaxation. In fact, there is not a priori justification to assume that many size  avalanches  will dominate the relaxation process within the all range of $j$ and $T$, a situation that was deeply analyze in references~\cite{Bonabeau_95,Bonabeau_96} and is discussed in a different context in section 5.

\section{Scaling relations}

Rather than to introduce directly the derivation of our scaling relations we prefer to start with a short review of some important scaling concepts of the theory of self organized criticallity. 

Following the first ideas of Bak and collaborators\cite{Bak_87,Bak_88,Bak_book}, those systems who behave as predicted
by SOC show a  distribution of avalanche sizes and times that follow power laws, i.e ,

\begin{equation}
P(s) \sim s^{-\tau}
\label{eq:P_so}
\end{equation}

\noindent and

\begin{equation}
P(t) \sim t^{-\tau_t}
\label{eq:P_to}
\end{equation}

\noindent respectively. For systems not exactly in the critical state, these expresions transform into:

\begin{equation}
P(s) \sim s^{-\tau}f(s/s_c)
\label{eq:P_s1}
\end{equation}

\noindent and

\begin{equation}
P(t) \sim t^{-\tau_t}f(t/t_c)
\label{eq:P_t1}
\end{equation}

\noindent where $s_c$ and $t_c$ reflect the departure of the system from criticality, $s_c \sim (j_c-j)^{-1/\sigma_1}$ and $t_c \sim (j_c-j)^{-1/\sigma_2}$ being $\sigma_1$ and $\sigma_2$ new critical exponents and 
where the function $f(x)$ has the following properties 
 $f(x)
\rightarrow cte$ if $x \rightarrow 0$ and $f(x) \rightarrow 0$ if $x \rightarrow \infty$ in order to recover the ``critical picture'' when $j \sim j_c$. 

In finite size systems $s_c$ and $t_c$ also reflect the effect of the sample dimensions through two new critical exponents $D$ and $z$. In fact, in analogy with the theory of critical phenomena $s_c \sim L^D$ and $t_c \sim L^z$. Furthermore, the coherence length $\xi$ diverges at the critical state as: 

\begin{equation}
\xi \sim (j_c-j)^{-\nu}
\label{eq:chi_j}
\end{equation}

From the definitions of $s_c$, $t_c$ and the equation (\ref{eq:chi_j}) it is straighforward to show that $s_c \sim \xi^{1/\nu\sigma_1} $ and $t_c \sim \xi^{1/\nu\sigma_2}$. Moreover, since for a finite size system at the critical state $\xi=L$, our first scaling relation takes the form:

\begin{equation}
\frac{D}{z}= \frac{\sigma_1}{\sigma_2}
\label{eq:scaling}
\end{equation}

As already discussed above, the integrated distributions of avalanche sizes and times, calculated in section 2, $D_{int}(s)$ and $D_{int}(t)$ result, since the system is relaxing, from  avalanches obtained for values of  current densities ranging from $j_c$ to  $j$. These distributions are different from those obtained in typical field driven experiments or simulations since the last are obtained ``in principle'' for a fixed value of current density $j_c$ which, indeed, determines the criticality of the system. 

Then, it is natural to assume that $D_{int}(s)$ and $D_{int}(t)$  are related to the distributions obtained just at the critical state, $D(s)$ and $D(t)$, by the following formulae:
 
\begin{equation}
D_{int}(s) \sim \int^{0}_{j_c} s^{-\tau} f(s/s_c) dj
\label{eq:P_s_f_s}
\end{equation}

\noindent and

\begin{equation}
D_{int}(t) \sim \int^{0}_{j_c} s^{-\tau_t} f(t/t_c) dj
\label{eq:P_t_f_t}
\end{equation}

\noindent which inmediately explain the meaning of the label ``integrated'' used for these distributions.

Then substituing the definitions of $s_c$ and $t_c$ in equations (\ref{eq:P_s_f_s}) and 
(\ref{eq:P_t_f_t}) and after a simple change of variables, we obtain the following expresions for the integrated distributions of avalanche sizes and times:

 \begin{equation}
D_{int}(s) = s^{-\tau+\sigma_1}
\int_0^{s(-j_c)^{1/\sigma_1}} \sigma_1 x^{\sigma_1-1} f(x) dx
\label{eq:P_int_s_j}
\end{equation}

\begin{equation}
D_{int}(t) = s^{-\tau_t+\sigma_2}
\int_0^{s(-j_c)^{1/\sigma_2}} \sigma_2 x^{\sigma_2-1} f(x) dx
\label{eq:P_int_t_j}
\end{equation}

\noindent which prove that for $s$ large enough both integrals are constants, and there is not a cut-off length in the integrated distributions, result already obtained in our simulations (see figures 3 and 4).

 Also from equations (\ref{eq:P_int_s_j}) and  (\ref{eq:P_int_t_j}) and the definitions of $D_{int}(s)$ and $D_{int}(t)$ we can inmediatly obtain the following scaling relations

\begin{equation}
\tau_n=\tau+\sigma_1
\label{eq:scal_1}
\end{equation}

\begin{equation}
\tau_{tn}=\tau_t+\sigma_2
\label{eq:scal_2}
\end{equation}

\noindent which in combination with (\ref{eq:scaling}) leads to:

\begin{equation}
\frac{D}{z}=\frac{\tau_n -\tau}{\tau_{tn}-\tau_t}
\label{eq:scal_f}
\end{equation}

In this way expresion (\ref{eq:scal_f}) establishes a connection between the exponents obtained in field driven experiments or simulations, $\tau,\tau_t,D,z$ and those from thermally activated avalanches $\tau_n,\tau_{tn}$. In fact, the results obtained in our simulations and those obtained in references \cite{Bassler_98, Cruz_99} hold the previous relation.

However, some points deserve further discussion. The power law divergence of $s_c$, $t_c$ and  $\xi$ are strictly valid close to the critical state, $j_c$.  Far away from this state these divergencies no longer exactly hold, however considering the good results obtained in the check of our calculations and the scaling law (\ref{eq:scal_f}) we believe that this last assumption is not relevant for the solution of the model. Also, the scaling law (\ref{eq:scal_f}) was obtained assuming the complete relaxation of the system, so it is difficult to be proved in real experiments.

\section{Applicability}

Our previous picture assumes that a thermally activated vortex jump would affect its neighborhood generating an
 unstability that leads to a cascade of vortex 
jumps related to the
vortex distribution into the sample.

However, it is well known the existence of a characteristic time 
for thermally activated phenomena $t_{th}=t_o \exp{(U(j)/kT)}$ 
representing the time a vortex  spends at a pinning site 
before jumping due to thermal 
activation \cite{Blatter_94}.

This means that our model will be valid if these avalanches occur within  times lower than $t_{th}$, i.e. the avalanches should develop fast enough to be mutually independent. This resembles the idea developed by Vespignani et al~\cite{Vespignani_98} in the context of sandpile and forest fire models. They showed through simulations and mean field considerations that one necessary condition for the occurrence of SOC, at least in these models, is the separation of time scales between the external exitation and the response of the system. 

Then, as mentioned above, the maximum time an avalanche persits is, $t_c
=t_{co}(1-j/j_c)^{-1/\sigma_2}$ where $t_{co}$ is the time a vortex spends moving from one site to another, and of course depends on the local  current and flux density in the system. Considering that the vortices are separated a distance $a$ the time they spend traveling this distance is

\begin{equation}
t_{co}=\frac{a}{v}
\label{eq:t_co}
\end{equation}

\noindent where $v$ depends on the Lorentz force acting on the vortex $v=j \Phi_o/\eta$, and $a=(\Phi_o/B)^{1/2}$ which inmediatly gives the following dependence of $t_{co}$ with $j$ and $B$.

\begin{equation}
t_{co}=\frac{\eta}{j\sqrt{(\Phi_{o}B)}}
\label{eq:t_co_j}
\end{equation}

In the critical state $B$ varies along the sample. This variation is, even in the presence of thermal activation, very well accounted by the Bean 
model~\cite{Bean_64,Abufalia_96}. This means that for a fully penetrated sample

\begin{equation}
B(x)=\mu_o H - \mu_o jx
\label{eq:B_x}
\end{equation}

\noindent where $H$  is the external field. Then, substituing equations (\ref{eq:t_co_j}) and (\ref{eq:B_x}) in the definition of $t_c$ we obtain that

\begin{equation} 
t_{c} = \frac{\eta (1-j/j_c)^{-1/ \sigma_2}}
{j \Phi^{1/2}_o \sqrt{\mu_o H - \mu_o jx}}
 \label{eq:tc_fin} 
\end{equation}

Now, to determine the regime of applicability of our model, we must verify under which conditions the inequality $t_{c}<<t_{th}$ holds.
 
From the experimental point of view, the relevant avalanches to be detected measuring 
 the fluctuations in the magnetization decay~\cite{Aegerter_98} are those starting at the border of the sample, since are those who produce changes in $m$. Moreover, the avalanches starting at the border of the sample are also those with larger duration times since they have a larger area for spreading, (remember that the critical state in a type-II superconductor is symmetric with respect to the center of the sample).
Then, we may assume  in (\ref{eq:tc_fin}) $x=0$ and obtain the following inequality:

\begin{equation} 
\frac{\eta (1-j/j_c)^{-1/ \sigma_2}}
{j \Phi^{1/2}_o \sqrt{\mu_o H}} << t_o exp(U(j)/kT)
 \label{eq:inequal_a} 
\end{equation} 

\noindent which can be written as:

\begin{equation} 
\frac{j^{*}}{j} \frac{1}{(1-j/j_c)^{1/\sigma_2}}
<< exp(U(j)/kT)
 \label{eq:inequal_b} 
\end{equation} 

\noindent where $j^{*}=\frac{\Phi_o \eta}{\sqrt H} t_o$. Assuming, for example $U(j)=U_o \ln(j_c/j)$~\cite{Blatter_94}, the previous inequality takes the form:

\begin{equation} 
\frac{j^{*}j^{\alpha-1}}{j^{\alpha}_c}{(1-j/j_c)^{-1/\sigma_2}}
<< 1
 \label{eq:inequal_c} 
\end{equation} 

\noindent $\alpha=U_o/kT$.

It is then straighforward to demostrate that equation (\ref{eq:inequal_c}) 
holds under the following conditions, if $\alpha>>1$  $j$ must be much lower than $j_c$ ($j<<j_c$). In the opposite case $j^{*}<<j<<j_c$. Similar expresions can be derived for the Anderson-Kim potentials and from potentials derived from the Collective Pinning Theory\cite{Blatter_94}.

These conditions are the consequence of the competition
 between the increase of
$t_{th}$ when $j \rightarrow 0$ and the divergence of $t_{c}$ when
$j \rightarrow 0$ and $j \rightarrow j_c$, see equation~(\ref{eq:tc_fin}), and 
can be interpreted in the following way.
  Close to $j_c$ the avalanche durations are  
very high because the avalanche sizes become huge, so one always 
need to be far from $j_c$, a situation 
often accounted in high temperature superconductors \cite{Malozemoff},
 to assure that $t_{av}<<t_{th}$.
Particularly for $U_o<<kT$, when thermally activated jumps
 become frequent 
($t_{th}$ small), 
high enough currents ($j>>j^{*}$) are also neccesary to assure 
rapid vortex motion during 
the avalanche, and hence low avalanche time durations. From 
the experimental point of view these conditions should be seen
 with some caution. For example,  
since $j_{c}$ decays with  temperature \cite{Blatter_94}, 
for $U_{o}<<kT$, the range of current densities where thermally
 activated avalanches could appear is still narrower than that 
suggested by
a simple inspection to the formula
$j^{*}<<j<<j_c$ so, we strongly recommend to look for these
 avalanches  at low temperatures and in very disordered 
systems were $U_o>>kT$.

In the light of these results it is useful to come back to the experiment of Aegerter~\cite{Aegerter_98}. He found one critical exponent characterizing the avalanche size distribution during the relaxation of the magnetization and that this exponent was independent of the temperature of the system. Neither of these  results contradict our model. Even when his critical exponent was 2.0 and our $\tau_n=2.7$, figure 5 indicates that small exponents are associated to small relaxation times in our model. This suggest that, if in the experiment of reference~\cite{Aegerter_98} the time window had been shifted to larger times, and exponent closer to our one would have been observed. This does not mean, of course, that such a shift can be trivially performed in the practice.

 In addition he found, during the relaxation, one initial regime where avalanches are not power law distributed. While he explained this due to a transient period the system takes to reach the SOC, our results suggest a different explanation. During this period the system is still too close to the critical state
$j \sim j_c$,  and power law avalanches are not yet developed since the thermally activated avalanches overlap each other. This 
 explanation is consistent with the long time associated to this transient period, and to the dependence of this time with the temperature. Experimentally Aegerter found that larger times are associated with larger temperatures and in fact, in our model larger temperatures imply the neccesity of lower values of the relation $j/j_c$ to found power 
law distributed avalanches and this means larger transient periods.

\section{Conclusions}

In conclusion, we developed a simple scenario to explain recently reported thermally activated avalanches power law distributed for type-II superconductors. 
We proved that the exponents associated with these distributions depend on the time interval of the measurement. We also proved that the exponents characterizing a distribution of thermally activated avalanches obtained during the whole relaxation experiment, (i.e, from the critical to the equilibrium states), are related to those obtained in field driven experiments by scaling relations, a situation also supported by our simulations.  The conditions for the appearance of these
 avalanches were discussed and it was also proved that, in a rough
 approximation, they do not  depend on the pinning mechanism in the
 sample. All ours theoretical predictions are consistent with the known experimental results.

\section*{Acknowledgments}
We are  very grateful to 
A. V\'azquez for many interesting discussions and suggestions.
We acknowledge also usseful 
 comments from M. Paczuski, K. Bassler, 
D. Dom\'{\i}nguez,  E. Osquiguil, O. Sotolongo and C. Rodr\'{\i}guez.
E. A. acknowledges partial financial support from the World Laboratory Center for Pan-American collaboration in Science and Technology, the Texas Center for Superconductivity, and the Department of Physics, University of Houston.

 \newpage \section*{Figure captions}

  \hspace{0.5cm}

{\bf Figure 1} Magnetic relaxation curves for systems 
of sizes  $L=60,100,200$. $U(j) \sim j_c/j$, $U_o/kT=\infty$. 
The inset shows the data  collapse of the curves. 

 {\bf Figure 2} Magnetic relaxation curves for systems 
of sizes $L=60,100,200$. $U(j) \sim  j_c/j$, $U_o/kT=10$
The inset shows the  collapse of the curves.  

{\bf Figure 3} Avalanche sizes distribution for $L=200$, 
$U(j) \sim (1-j/j_c)$, $U_o/kT=10$.  

{\bf Figure 4} Avalanche times distribution for $L=200$, 
$U(j) \sim (1-j/j_c)$, $U_o/kT=10$.  

{\bf Figure 5} Avalanche size distribution for $L=200$,
$U(j) \sim j_c/j$, $U_o/kT=10$. From the upper to the lower curve:
 $t=1-10$, $t=11-100$, $t=101-1000$, $t=1001-10000$ and $t=10001-100000$ m.c.s. The straigh line represents a power law with exponent 2.7.

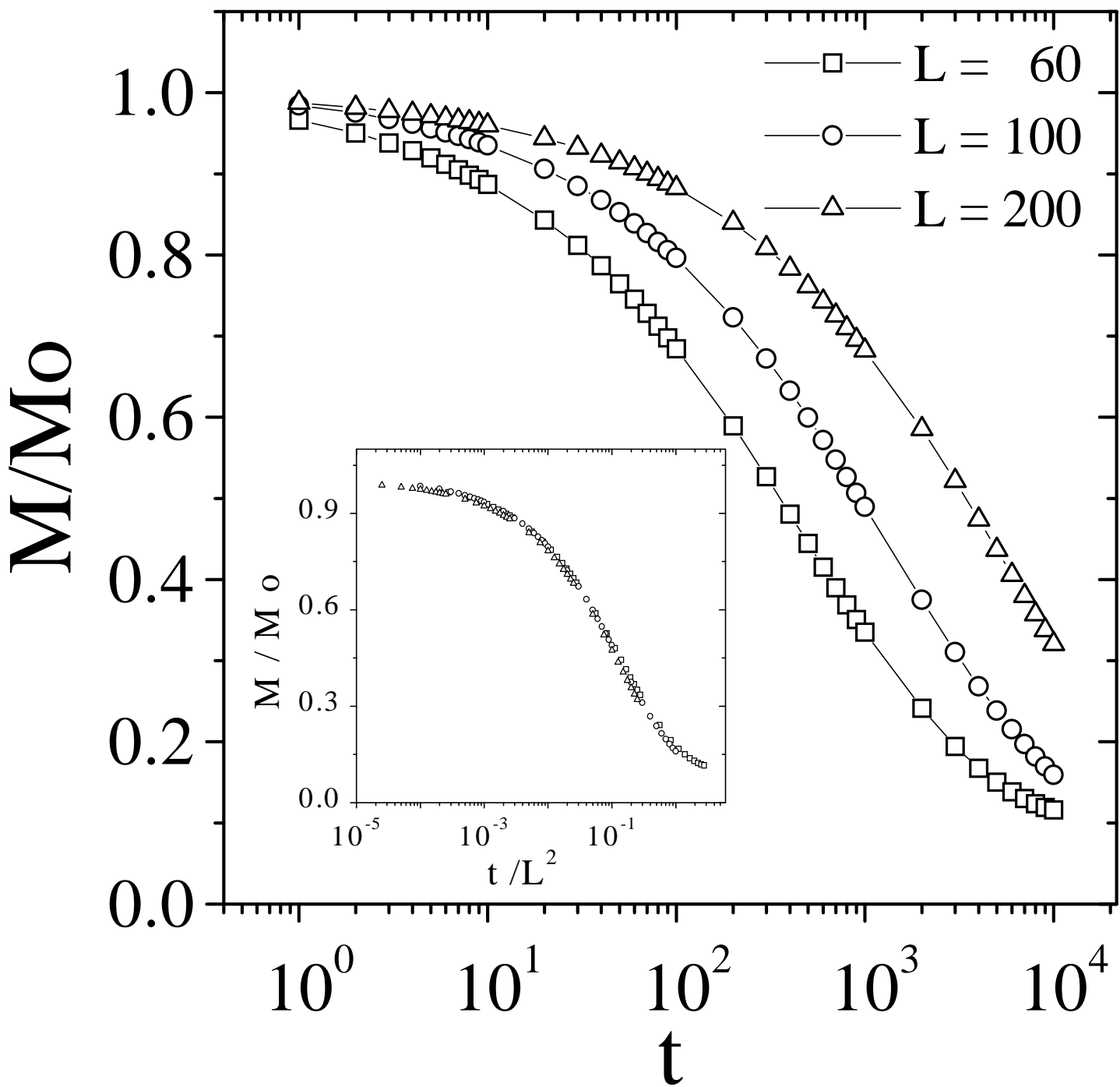
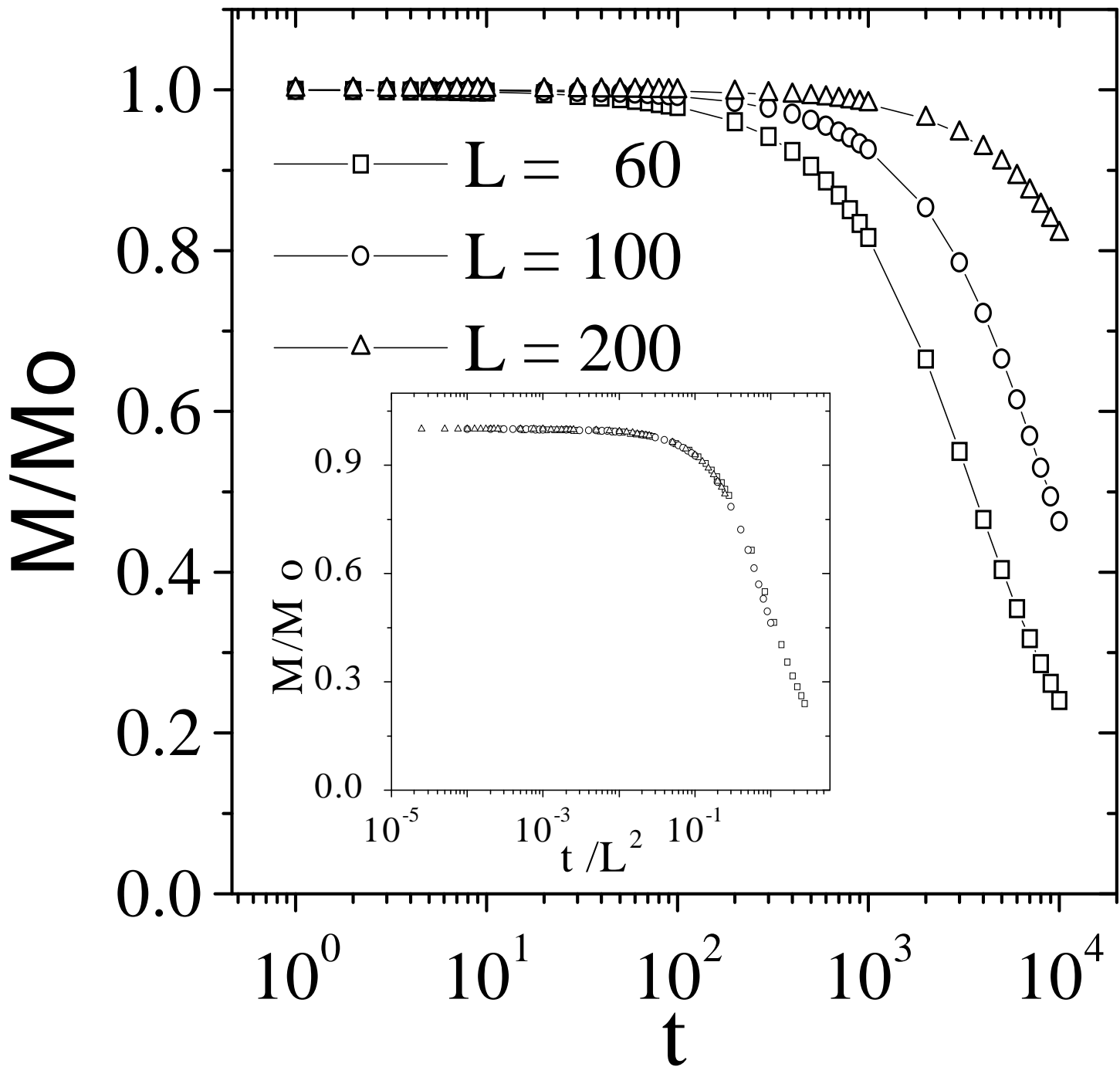
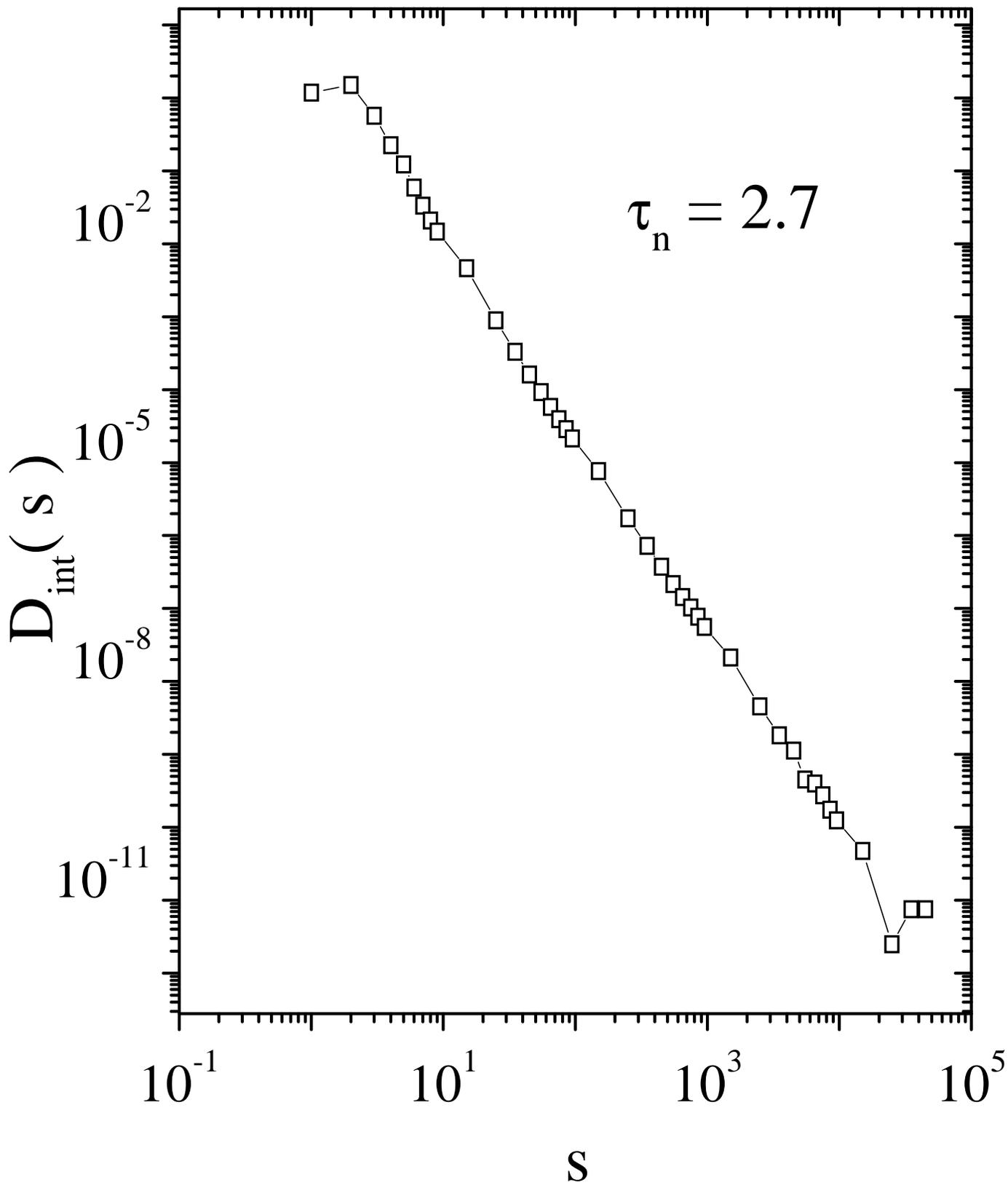
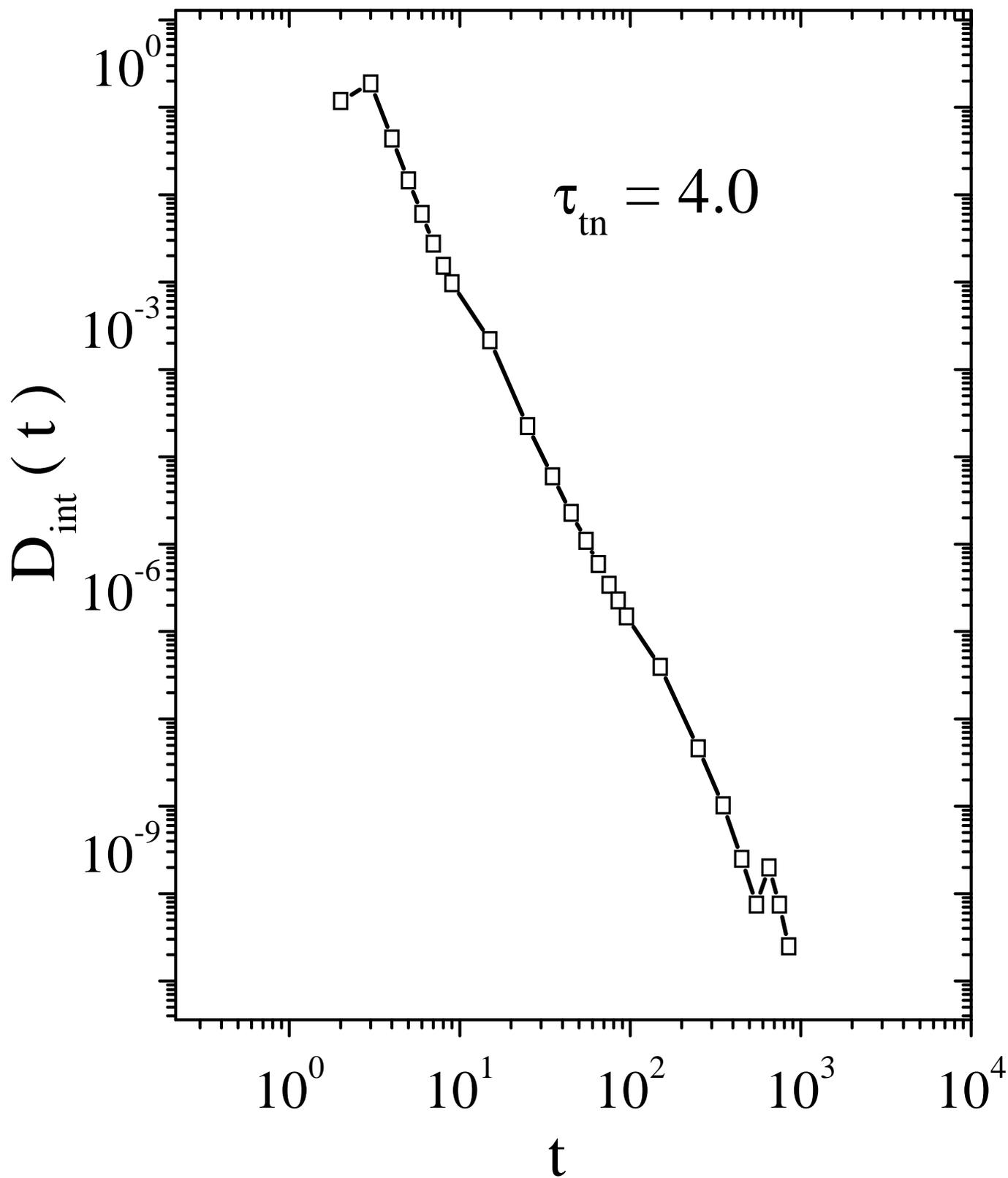
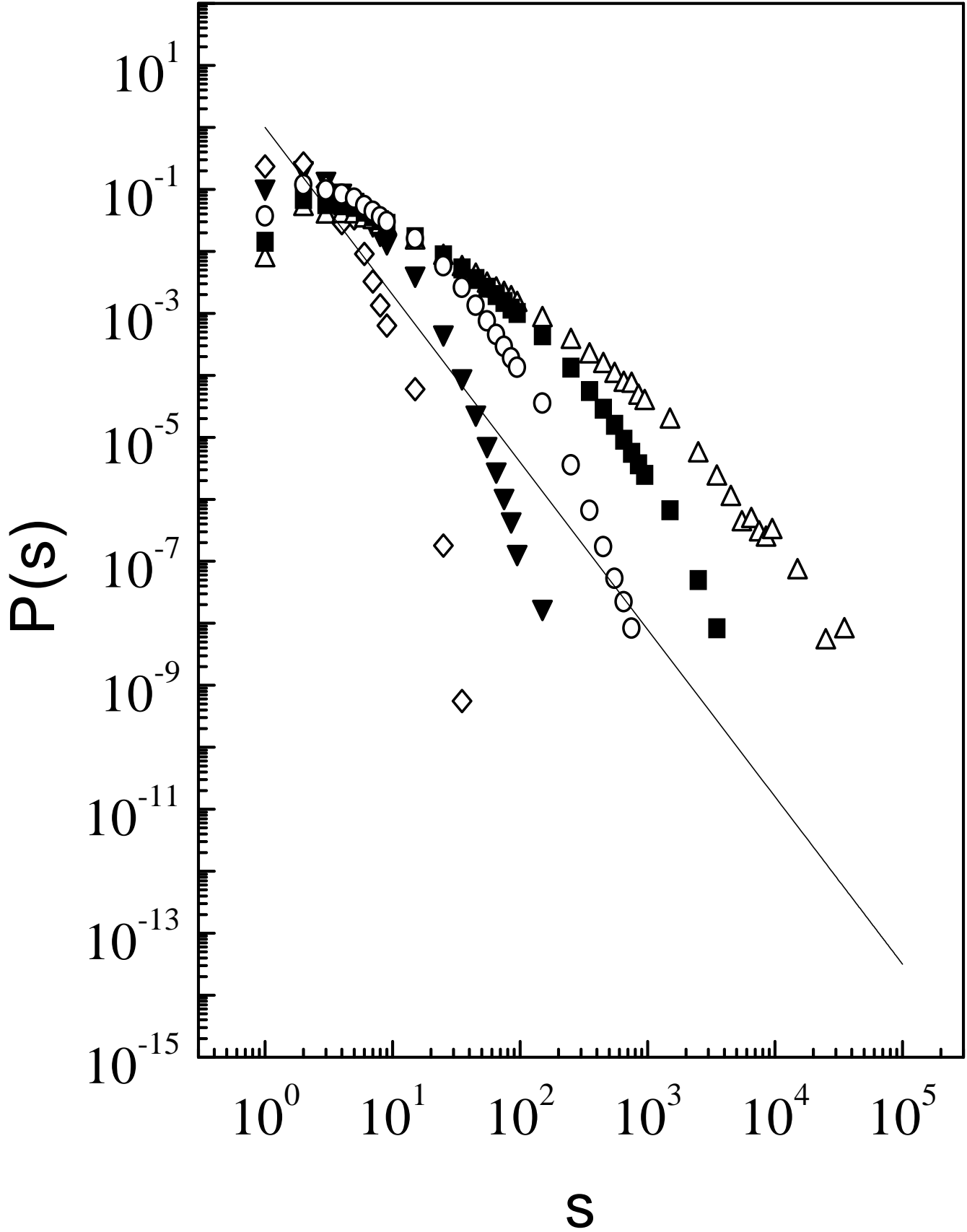
\end{document}